\documentclass[conference]{IEEEtran}
\IEEEoverridecommandlockouts
\usepackage{cite}
\usepackage{amsmath,amssymb,amsfonts}
\usepackage{algorithmic}
\usepackage{graphicx}
\usepackage{textcomp}
\usepackage{caption}
\usepackage{tabularx}
\usepackage{subcaption}
\usepackage{hyperref}
\usepackage{framed}
\usepackage{flushend}
\usepackage{booktabs}
\usepackage{xcolor}
\def\BibTeX{{\rm B\kern-.05em{\sc i\kern-.025em b}\kern-.08em
    T\kern-.1667em\lower.7ex\hbox{E}\kern-.125emX}}

\usepackage{tcolorbox}
\tcbuselibrary{theorems}

\newtcbtheorem[number within=section]{definition}{Definition}%
{colback=gray!5,colframe=gray!10!gray,fonttitle=\bfseries}{th}

\usepackage[strict]{changepage}
\definecolor{formalshade}{gray}{0.96}
\definecolor{darkblue}{RGB}{0,80,155}
\definecolor{green}{RGB}{177,201,31}

\newenvironment{finding}{%
	\small
	\MakeFramed{\advance\hsize-\width\FrameRestore}%
	\noindent\hspace{-4.55pt}
	\begin{adjustwidth}{}{7pt}%
		\vspace{1pt}\vspace{1pt}%
	}
	{%
		\vspace{2pt}\end{adjustwidth}\endMakeFramed%
}

\newenvironment{mydefinition}{%
	\small
	\MakeFramed{\advance\hsize-\width\FrameRestore}%
	\noindent\hspace{-4.55pt}
	\begin{adjustwidth}{}{7pt}%
		\vspace{1pt}\vspace{1pt}%
	}
	{%
		\vspace{2pt}\end{adjustwidth}\endMakeFramed%
}

\begin{document}

\title{Towards Shaping the Software Lifecycle \\with Methods and Practices}

\author{\IEEEauthorblockN{
	Jil Kl\"under\IEEEauthorrefmark{1},
	Melanie Busch\IEEEauthorrefmark{1}, 
	Natalie Dehn\IEEEauthorrefmark{1},
	Oliver Karras\IEEEauthorrefmark{1}
}
	\IEEEauthorblockA{\IEEEauthorrefmark{1}Leibniz University Hannover\\ Software Engineering Group \\Hannover, Germany \\Email: \{jil.kluender, melanie.busch, oliver.karras\}@inf.uni-hannover.de, natalie.dehn@live.de
	}
}

\maketitle

\begin{abstract}
As software projects are very diverse, each software development process must be adjusted to the needs of the project and the corresponding development team. Frequently, we find different methods and practices combined in a so-called hybrid development method. Research has shown that these hybrid methods evolve over time and are devised based on experience. However, when devising a hybrid method, the methods and practices used should cover the whole software project with its different phases including, among others, project management, requirements analysis, quality management, risk management, and implementation.
In this paper, we analyze which methods and practices are used in which phase of a software project. Based on an initial survey with 27 practitioners, we provide a mapping of methods and practices to different project phases and vice versa. Despite the preliminary nature of our study and the small sample size, we observe three remarkable aspects: (1) there are discrepancies between the intended use of methods and practices according to literature and the real use in practice, (2) practices are used more consistently than methods, and (3) parts of the software lifecycle such as maintenance and evolution are hardly covered by widely distributed methods and practices. Consequently, when devising a development process, it is worth a thought whether all phases of the software lifecycle are addressed or not. 
\end{abstract}

\begin{IEEEkeywords}
Hybrid method, software lifecycle, software process, method, practice, selection, survey study
\end{IEEEkeywords}

\section{Introduction}
A process that is suitable to carry out a software project usually consists of several phases going far beyond the implementation of the software~\cite{royce1998software}. These phases include, among others, project management, requirements analysis and management, quality assurance, and---of course---the technical phases such as implementation, testing, and maintenance~\cite{ruparelia2010software}. A holistic development process should address all these relevant phases of the software lifecycle, which we also refer to as \textit{project disciplines}. For example, consider a process based on Scrum with its practices (such as daily stand-up meetings, retrospectives, backlog management, and an onsite-customer) as a project management framework (also partially covering the requirements engineering). This process is extended with some eXtreme programming practices such as pair programming (for quality management), test-driven development to structure the implementation and testing, as well as CI/CD for the integration. Despite the number of methods and practices combined in this process, several further relevant project disciplines such as maintenance, configuration management, and risk management are still not addressed.

The combination of different (agile or plan-driven) methods and practices---a so-called hybrid development method~\cite{klunder2019catching}---as in the example described before is state-of-the-practice~\cite{tell2019hybrid} with three out of four companies using such a hybrid method~\cite{klunder2019catching}. However, it is not yet completely clear, how suitable methods and practices can be identified to form such a hybrid method~\cite{tell2019hybrid}. According to the large-scale international HELENA study~\cite{Kuhrmann:2018aa}, 30\% of the projects select the methods and practices on demand. Other processes are based on defined rules (17\%), tailored by the project manager (16\%), or according to customer demands (14\%)~\cite{Kuhrmann:2018aa}. Almost 80\% of these development processes evolve over time based on learning from past projects~\cite{klunder2019catching}. In addition, when devising such a hybrid method, several context factors have to be considered, including \textit{situational factors}~\cite{clarke2012situational} and \textit{tailoring criteria}~\cite{kalus2013criteria}. Research provides first attempts to support the creation of a (hybrid) development method. For instance, Tell et al.~\cite{tell2020towards} present a statistical construction method, and Klünder et al.~\cite{klunder2020determining} provide an overview of context factors that influence the use of methods and practices. Both approaches support the construction of a hybrid method, as soon as a starting point, i.e., a pre-selection of methods and/or practices, is defined. In this case, both approaches provide recommendations with which methods and practices the pre-selected ones can be combined~\cite{klunder2020determining, tell2020towards}.

\subsubsection*{Problem Statement} Tell et al.~\cite{tell2019hybrid} present a ``typical'' hybrid method consisting of a subset of the methods \textit{Scrum, Iterative Development, Kanban, DevOps,} the \textit{Classic Waterfall Process, eXtreme Programming}, and \textit{Lean Software Development}. These methods are extended by a variety of practices. However, it appears that even a selection of a huge amount of methods and practices does not adequately cover the whole software lifecycle. Nevertheless, a software process needs to address all the relevant phases to reduce the risk of project failure (for example due to a missing risk management). That is, if applicable for the respective project, a hybrid method needs to contain methods and/or practices for the different project disciplines, i.e., for the requirements analysis, the implementation, testing, maintenance, etc.  

\subsubsection*{Objective} In this paper, we aim to investigate whether development processes used in practice cover the whole software lifecycle. In particular, we analyze which methods and practices are used in which phase of a project. For example, we investigate whether Scrum is used for the project management only, or also for other disciplines. This way, we can uncover phases of the software lifecycle that can be hardly covered by using standard methods and practices only (and, hence, require more attention when devising a development process), and we can point to the (mis-)\-application of methods and practices throughout the process.  

\subsubsection*{Contribution} Based on a pilot study with 27 practitioners reporting on in total 1,257 applications of methods and practices in different project disciplines, we analyze which methods and practices are used in which phase of the software lifecycle. We report on the use of \textit{Scrum}, \textit{DevOps}, and \textit{Kanban} as top-3 methods, as well as on the use of \textit{Code Reviews}, \textit{Daily Stand-up Meetings}, and \textit{Coding Standards} as top-3 practices. Our initial results show that these methods and practices are used for almost all project disciplines covered by our pilot study. Furthermore, looking at project disciplines mentioned more than 100 times in our study, we observe a wide variety in the methods and practices used to support \textit{Implementation and Coding}, \textit{Project Management}, \textit{Integration and Testing}, and \textit{Quality Management}. 

\subsubsection*{Outline}
The rest of the paper is structured as follows: In Section~\ref{sec:related-work}, we present related work. Section~\ref{sec:research-design} presents our research design including the research goal, the data collection, data analysis and validity procedures. In Section~\ref{sec:results}, we present our results that we discuss in Section~\ref{sec:discussion}. We conclude the paper in Section~\ref{sec:conclusion}. 

\section{Related Work}\label{sec:related-work}
Several researchers studied the use of methods and practices in the development process. 
In the HELENA study, Kuhrmann et al.~\cite{kuhrmann2017hybrid,Kuhrmann:2018aa} analyze which methods and practices are used in software systems development. According to their results, so-called hybrid methods---combinations of different methods and practices---are state-of-the-practice. These results are supported by the findings of Vijayarathy and Butler~\cite{vijayasarathy2015choice}. They focused on software development processes in practice. Based on their results, 45,3\% of the development approaches covered in their survey are hybrid, 33,1\% agile, 13,8\% traditional and 7,7\% iterative. Tell et al.~\cite{tell2019hybrid,tell2020towards} analyze such hybrid methods in detail by providing an overview of what a hybrid method is made of~\cite{tell2019hybrid}, and how it can be statistically constructed~\cite{tell2020towards}. Klünder et al.~\cite{klunder2020determining} investigate which context factors influence the choice of methods and practices. 

Besides the HELENA study, further studies analyze which methods and practices are used in practice, such as the State of Agile study~\cite{VersionOne:2006ol} and Status Quo Agile~\cite{Quo}.

For some project disciplines, researchers provide a set of methods and practices that are established. In the field of requirements engineering, Méndez Fernández et al.~\cite{fernandez2017naming} founded the initiative \textit{Naming the Pain in Requirements Engineering (NaPiRE)}: They conduct an international distributed family of surveys every two years, which, among other things, asks the participants about occurring problems in Requirements Engineering in practice and the status quo. In addition to examining the problems solely, they also try to identify patterns between the problems and contextual factors. Wagner et al.~\cite{wagner2019status} present an empirical theory of Requirements Engineering practices in order to provide assistance to software engineers in choosing appropriate Requirements Engineering methods and techniques. 
Kassab et al.~\cite{kassab2014state} focus on the current status of practice in requirements engineering regarding software development practices and techniques in their research. They conducted an online survey with practitioners. Their results show, among other things, that the three techniques most frequently chosen for the elicitation phase are brainstorming, interviews, and user stories.

Related works show a focus on Requirements Engineering, but to the best of our knowledge, there are no comparable studies for all phases in the software lifecycle. Within the scope of our work, we provide a broader overview by looking at methods and practices as well as at their use in specific project disciplines. In future work, these results should be provided in more detail for other project disciplines. 

\section{Research Design}\label{sec:research-design}
To reach our overall research goal of \textit{analyzing the methods and practices used in the different project disciplines}, we conducted a survey study with practitioners. In the following, we present the research questions, the used survey instrument, the data collection, and the data analysis. 

\subsection{Research Goal and Research Questions}\label{sec:rqs}
The overall objective of the research presented in this paper is to \textit{analyze which methods and practices integrated in the overall development process of a team are used in the different project disciplines}. For this purpose, we aim to answer the following research questions:\\
\textbf{RQ1: }\textit{ Which methods and practices are used by the companies that participated in our study?} Before studying which methods and practices are used in the respective disciplines, we analyze which of them are used in general in our pilot study. Besides, answering this question provides insights on the reliability of our data. Specifically, we investigate whether the sample is somewhat representative for software producing companies or covers exceptional cases only. The answer to this question defines the baseline for the next question. \\
\textbf{RQ2: }\textit{ For which project disciplines are the methods and practices used?} We analyze for which project disciplines the methods and practices are used by a majority of the participants to get an overview of the different use cases of a specific method or practice.\\
\textbf{RQ3: }\textit{ Which methods and practices are used to support different project disciplines?} By answering this research question, we reverse the analysis for RQ2. For the project disciplines, we analyze which methods and practices are primarily and secondarily used. We investigate the variety in the method and practice use along the software lifecycle.

\subsection{Survey Design}
We collected our data using the survey method~\cite{robson2016real}. We developed an online questionnaire to collect data from software projects. The unit of analysis was an ongoing or finished software project. 

\paragraph{Development of the Survey Instrument} We developed the survey instrument in two iterations based on expert reviews by two independent researchers. The iterations led to slight adjustments of the survey instrument. For example, as a result of the second iteration, we added a question on the location of the company. The survey was made available in English, German, and Spanish, and consisted of six parts: (1)~Demographics, (2)~Method use, (3)~Mapping of methods to project disciplines, (4)~Practice use, (5)~Mapping of practices to project disciplines, and (6)~Closing. The questionnaire is available online \cite{kluender2021-dataset}. As the basis for the sets of methods and practices for the parts (2) and (4), we used the HELENA study~\cite{Kuhrmann:2018aa}. In the respective parts of the HELENA study, we find 24 methods and 36 practices which we integrated in our study. For each of the methods and practices, the survey participants were asked to state whether the method or practice is used in the process their answers refer to. In the next part, they stated for which project discipline the method or practice is used, e.g., whether it is mainly used for project management or for implementation. They were able to order the twelve given disciplines as to also mention secondary or tertiary use cases of a method or practice. For the project disciplines, our set of answers is also based on the HELENA study~\cite{klunder2019catching,tell2019hybrid}. However, we slightly adjusted the project disciplines to gain more fine-grained results: We divided the discipline \textit{Requirements Engineering} into \textit{Requirements Analysis} and \textit{Requirements Management} according to Börger et al.~\cite{Boerger.1999}. 

\paragraph{Data Collection} The data collection period was June to October 2020 following a \textit{convenience sampling strategy}~\cite{robson2016real}. As part of a bachelor thesis~\cite{dehn2020}, approximately 250 software-producing companies were invited to participate in the study. However, given a very small response rate of less then 10\%, we also invited practitioners from our personal networks via e-mail. In addition, we promoted the survey via social networks such as Twitter, LinkedIn, Xing, and GitHub. 

\subsection{Data Analysis Procedures}\label{sec:data-analysis-procedures}
The data analysis consisted of several steps that are described in detail in the following. We started with the full data set that is available online \cite{kluender2021-dataset}. 

\paragraph{Data Cleaning and Data Reduction}\label{sec:datacleaning} The first step was the data cleaning. As we only added a single question about the location of the company in the second version of the study compared to the first version, we combined the data points into one dataset and analyzed it as one. Therefore, we started with a dataset consisting of 57 data points. We  did not generally remove incomplete answers, as each combination of a method/practice and a project discipline was helpful for our analysis. That is, a data point is valuable for our analysis as soon at it contains information on a method or a practice and in which project disciplines it is used. Consequently, we removed data points that did not choose a single method/practice and the respective project discipline(s) in which it is used as these data points cannot contribute to the research presented in this paper. This reduction led to a dataset consisting of 28 data points. In addition, we removed one additional data point where the participant explicitly stated in the field for additional comments that the respondent did not prioritize the answers and always aligned all project disciplines in the given ordering to each method and practice. Therefore, this data point would have introduced a bias if included in the analysis. In this case, all methods and practices used are reported to be primarily used for \textit{Project Management} which was the project discipline mentioned first. These data cleaning and data reduction strategies led to a total number of 27 data points used for the analysis. 

\paragraph{Analyzing the method and practice use} To answer the first research question, we analyzed the method and practice use for the 27 data points. We descriptively studied how often the methods and practices are used by the participants. 

In addition, for each method and practice used by at least ten participants, we analyze for which project discipline this method/practice is used. Due to space limitations and limited reliability of the conclusions drawn on a smaller number of data points, we decided to exclude these methods and practices from further analysis. 

\paragraph{Used methods and practices per project discipline} For each project discipline, we calculate how often a method/practice is selected to be primarily or secondarily used by the participants. We define the terms \textit{primary use} and \textit{secondary use} as follows:

\begin{mydefinition}
	{\small
	\textbf{Definition 1 (Primary Use): }The \textit{primary use} $U^1(\mathcal{X})$ of a method or a practice $\mathcal{X}$ is the project discipline $\mathcal{D}$ for which the method or practice $\mathcal{X}$ is reported to be used the most (i.e., ranked on first place). We distinguish between the primary use $U^1 (\mathcal{X}, p_i)$ reported by a specific participant $p_i$, where $i \in I$ and $I$ is an index set, and the overall primary use of $U^1 (\mathcal{X})$ reported by the majority of the participants of our study. \\
	Consequently, the primary use of a method or practice describes for which project discipline it is \textbf{mainly} used.}
\end{mydefinition}

\begin{mydefinition}
	{\textbf{Definition 2 (Secondary Use): }\small The \textit{secondary use} $U^2(\mathcal{X})$ of a method or a practice $\mathcal{X}$ is the set of project disciplines $\mathcal{D}$ for which the method or practice $\mathcal{X}$ is reported to be used, but not primarily (i.e., not ranked on first place, but still mentioned). \\
	Consequently, the secondary use of a method or practice describes which project disciplines are addressed by the method/practice, but rather incidentally than planned.	}
\end{mydefinition}

\begin{figure*}[tbp]
	\centering
	\includegraphics[width=1\linewidth]{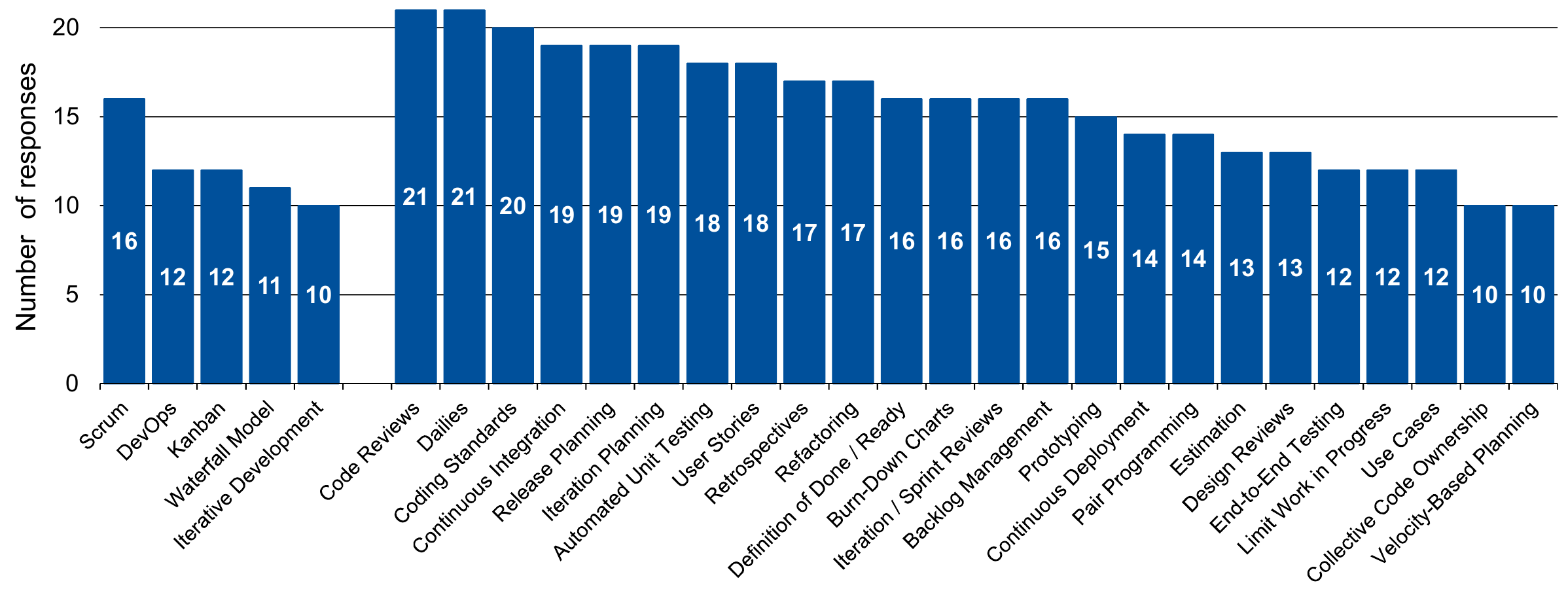}
	\caption{Total use of methods and practices (n=27). We only present methods and practices mentioned at least ten times. Participants were able to select multiple methods/practices.}
	\label{fig:total-use}
\end{figure*}
This selection indicates that the method/practice was used as part of the project discipline, for example if \textit{Project Management} is checked in conjunction with \textit{Scrum}, we can conclude that \textit{Scrum} is used for \textit{Project Management}. We distinguish between methods and practices that are primarily used to support the respective discipline (ranked on the first place; $U^1(\mathcal{X})$), and those that are not, i.e., that are primarily used for another part of the development process, and only affect the respective discipline secondarily ($U^2(\mathcal{X})$). 

\subsection{Validity Procedures}\label{sec:validity} To improve the validity of our results, we implemented different procedures affecting the replicability and to reduce bias. First, several researchers have been involved in the construction and the testing of the questionnaire. While one researcher set up the questionnaire based on the HELENA study~\cite{Kuhrmann:2018aa}, two internal researchers reviewed the process. Before publication, two independent researchers also thoroughly reviewed the questionnaire. 

The data analysis was mainly performed by two researchers and thoroughly reviewed by the other authors of this paper for quality assurance. 

We used the convenience sampling strategy to distribute the questionnaire~\cite{robson2016real}. This way, we lost the full control in terms of sampling, response rate, and the suitability of the participants. However, we are confident that answering the questionnaire was not possible without belonging to the target group of participants. The one case in which the questionnaire was not adequately answered was removed before analyzing the data (see Section~\ref{sec:data-analysis-procedures}). 

\section{Results}\label{sec:results}
We present the results focusing on the research questions posed in Section~\ref{sec:rqs} in the subsequent sections. 

\subsection{Demographics}
We performed the data cleaning as described in Section~\ref{sec:data-analysis-procedures}. The remaining dataset consisted of 27 data points. Each data point represents the answer of one respondent and reports on the methods and practices used, as well as the connection to the different project disciplines. 

Seven responses are related to very large companies with more than 2500 full time equivalents (FTEs), and three responses are related to large companies ({251-2499} FTEs). Nine respondents reported on working in a medium-sized company with {51-250} FTEs, and we retrieved five responses from small companies with {11-50} FTEs. Very small companies with a maximum of ten FTEs are not covered in our pilot study. We retrieved data points from almost all continents\footnote{Note that this multiple choice question was added as part of the questionnaire improvement which is why we only retrieved 15 answers to this question.}. Most of the companies are located in Europe (14 responses), followed by Asia (5 responses), and North America (4 responses). Three respondents reported on a company located in Africa, Australia, or South America (3 responses each). Two respondents reported on a company distributed on all continents. 

The top-3 business areas are the development of custom software (22 responses), IT consulting, training and services (13 responses), and the development of standard software (12 responses). The top-3 target application domains are web applications (13 responses), cloud applications and services (11 responses), and mobile applications (11 responses). 

\subsection{Method and Practice Use}
In a next step, we analyzed which methods and practices are used by the participants of our study. Figure~\ref{fig:total-use} shows how often each method and practice is used by the respondents. Note that we only consider methods and practices that are used by at least ten participants of the pilot study due to the limited informative value and the poor reliability of rarely used methods and practices. 

\subsubsection{Overview of Methods and Practices Used}
As evident from Figure~\ref{fig:total-use}, five methods are used by more than ten participants: \textit{Scrum} (16 responses), \textit{DevOps} and \textit{Kanban} (each 12 responses), the \textit{Classic Waterfall Model} (11 responses), and \textit{Iterative Development} (10 responses). For the other methods, we only observe a rare use (in terms of mentions by participants) with seven methods that are not used at all, and three methods only used by one company in our study. Due to the limited informative value for methods that are only rarely reported to be used, we decided to exclude these methods from further analysis. Therefore, in the rest of the analysis, we concentrate on \textit{Scrum, DevOps, Kanban,} the \textit{Classic Waterfall Model,} and \textit{Iterative Development}. These top-5 methods in our pilot study coincide with the top-5 used methods in the HELENA study (in a different order)~\cite{Kuhrmann:2018aa}. 

When looking at the practices, we observe a larger distribution. From the 36 practices covered by our study, 24 are used by more than ten participants. The most frequently used practices are \textit{Code Reviews} and \textit{Daily Stand-up Meetings} (21 responses each), and \textit{Coding Standards} (20 responses). 

\begin{finding}
	\noindent
	\textbf{Finding 1a:} The top-3 used methods are Scrum, DevOps, and Kanban. \\
	\textbf{Finding 1b:} The top-3 used practices are Code Reviews, Daily Stand-up Meetings, and Coding Standards.\\
	\textbf{Finding 2:} Practices are used way more frequently than methods. Five (out of 24) methods and 24 (out of 36) practices are used by at least ten participants.
\end{finding} 

\begin{figure*}[!t]
	\centering
	\begin{subfigure}[b]{0.49\textwidth}
		\centering
		\includegraphics[width=\textwidth]{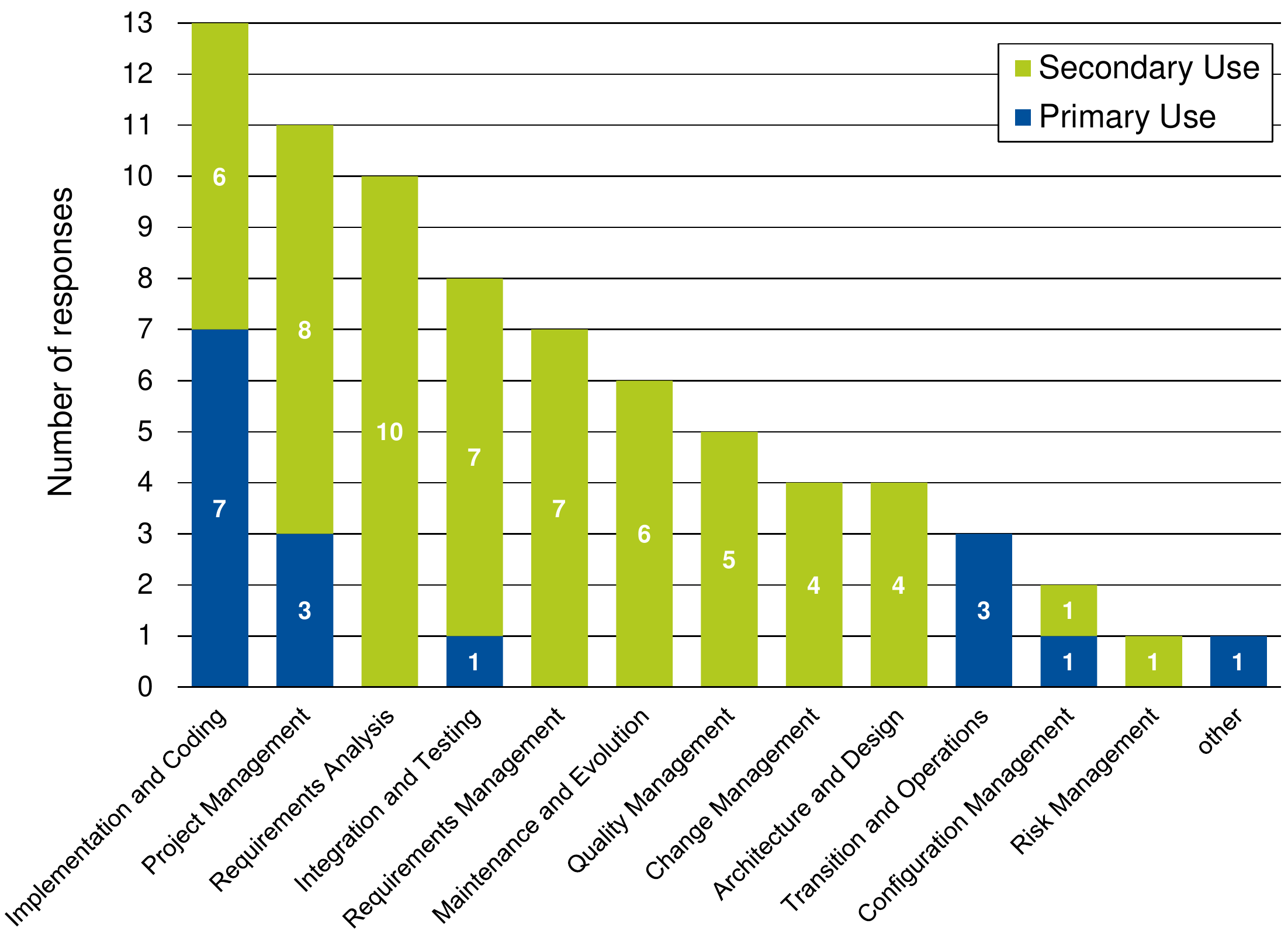}
		\caption{Use of Scrum per project discipline (n=16)}
		\label{fig:scrum}
	\end{subfigure}
	\hfill
		\begin{subfigure}[b]{0.49\textwidth}
		\centering
		\includegraphics[width=\textwidth]{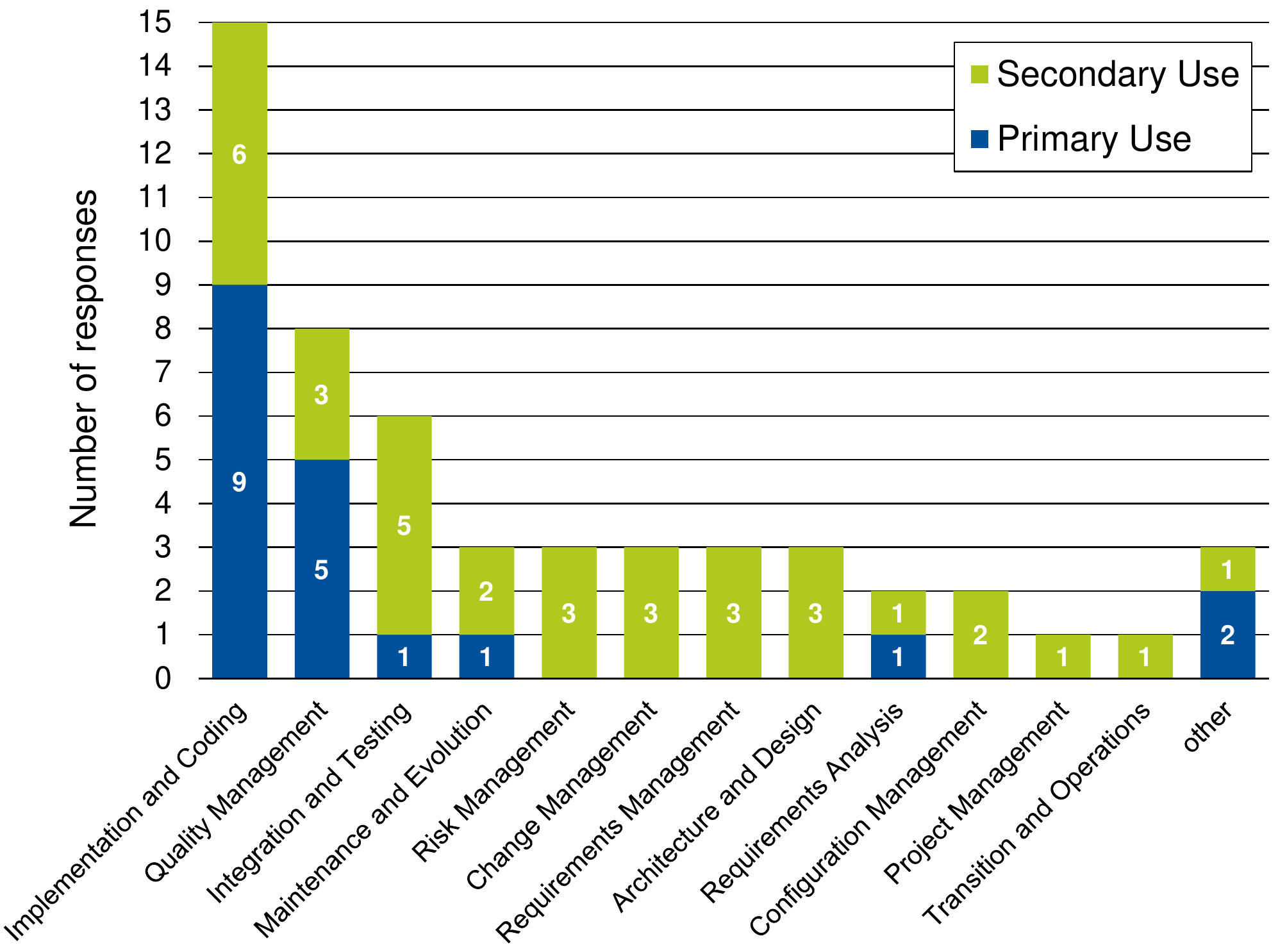}
		\caption{Use of Code Reviews per project discipline (n=21)}
		\label{fig:codereviews}
	\end{subfigure}
\hfill
	\begin{subfigure}[b]{0.49\textwidth}
		\centering
		\includegraphics[width=\textwidth]{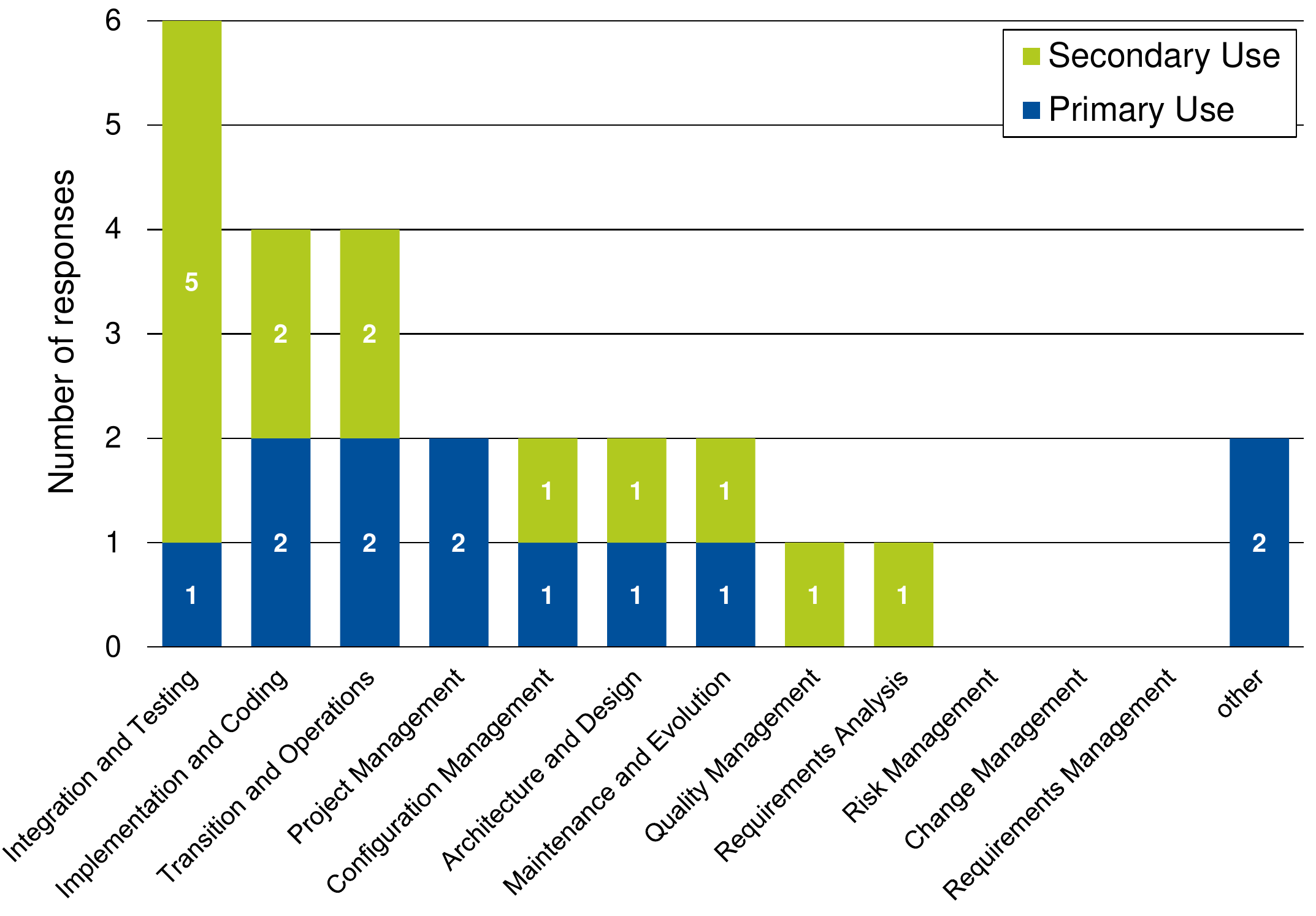}
		\caption{Use of DevOps per project discipline (n=12)}
		\label{fig:devops}
	\end{subfigure}
	\hfill
	\begin{subfigure}[b]{0.49\textwidth}
		\centering
		\includegraphics[width=\textwidth]{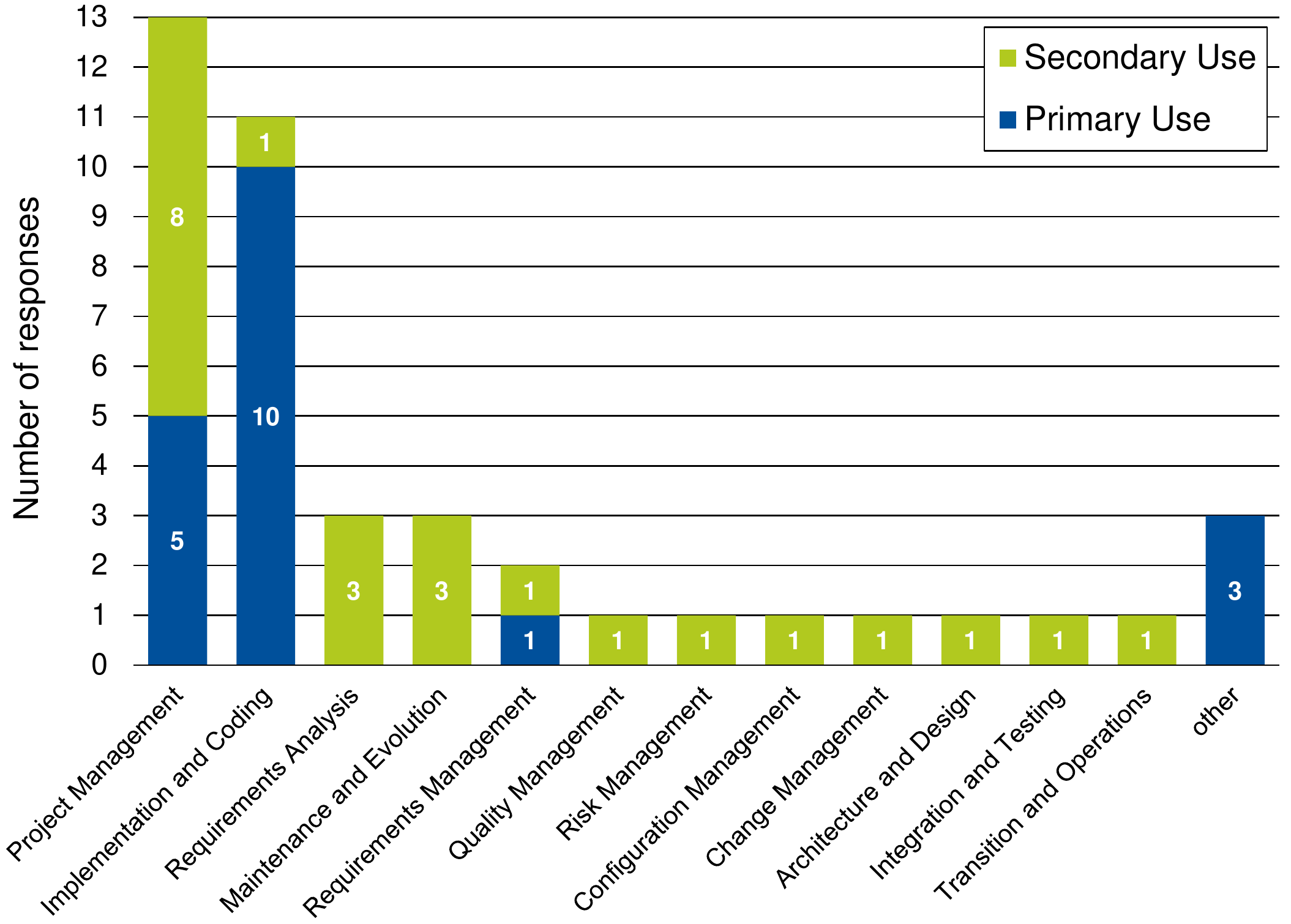}
		\caption{Use of Daily Stand-up Meetings per project discipline (n=21)}
		\label{fig:dailies}
	\end{subfigure}
	\hfill
	\begin{subfigure}[b]{0.49\textwidth}
		\centering
		\includegraphics[width=\textwidth]{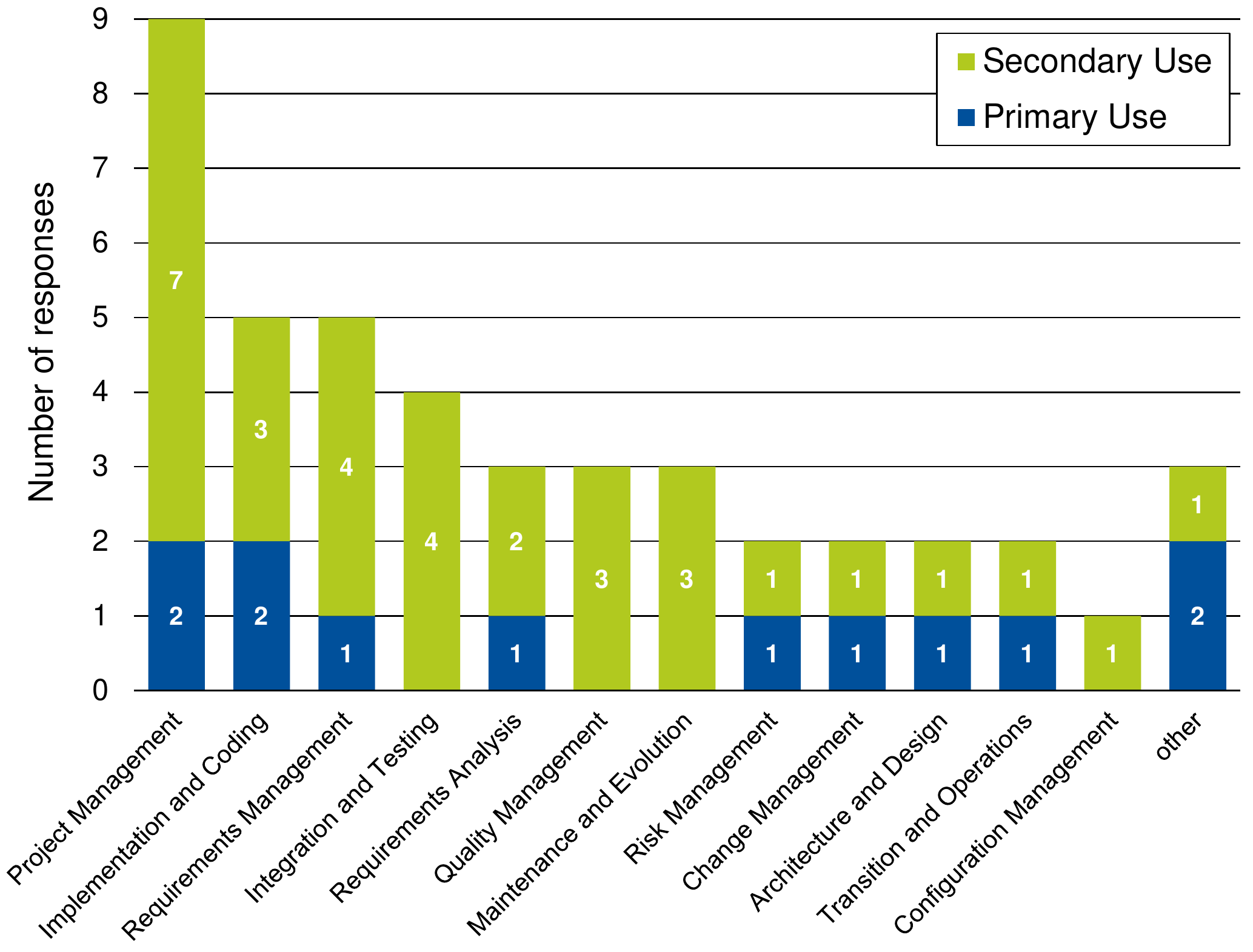}
		\caption{Use of Kanban per project discipline (n=12)}
		\label{fig:kanban}
	\end{subfigure}
	\hfill
	\begin{subfigure}[b]{0.49\textwidth}
		\centering
		\includegraphics[width=\textwidth]{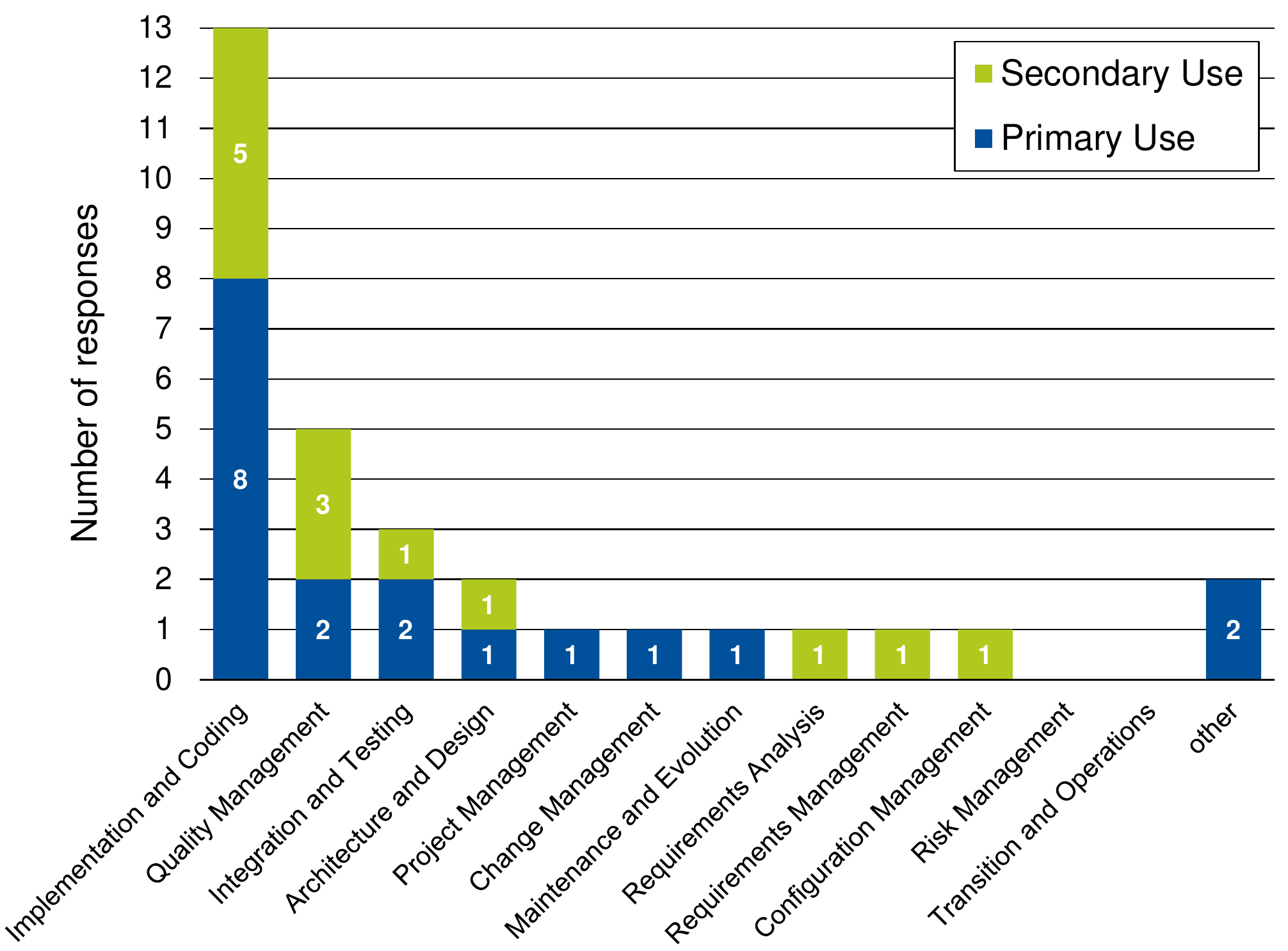}
		\caption{Use of Coding Standards per project discipline (n=20)}
		\label{fig:codingstandards}
	\end{subfigure}
	\hfill
	\caption{Use of the top-3 methods Scrum (a), DevOps (c), and Kanban (e), as well as the use of the top-3 practices Code Reviews (b), Daily Stand-up Meetings (d), and Coding Standards (f) in different project disciplines}
	\label{fig:methoduse}
	\label{fig:practiceuse}
\end{figure*}

\begin{figure}[t]
	\centering
	\includegraphics[width=1\columnwidth]{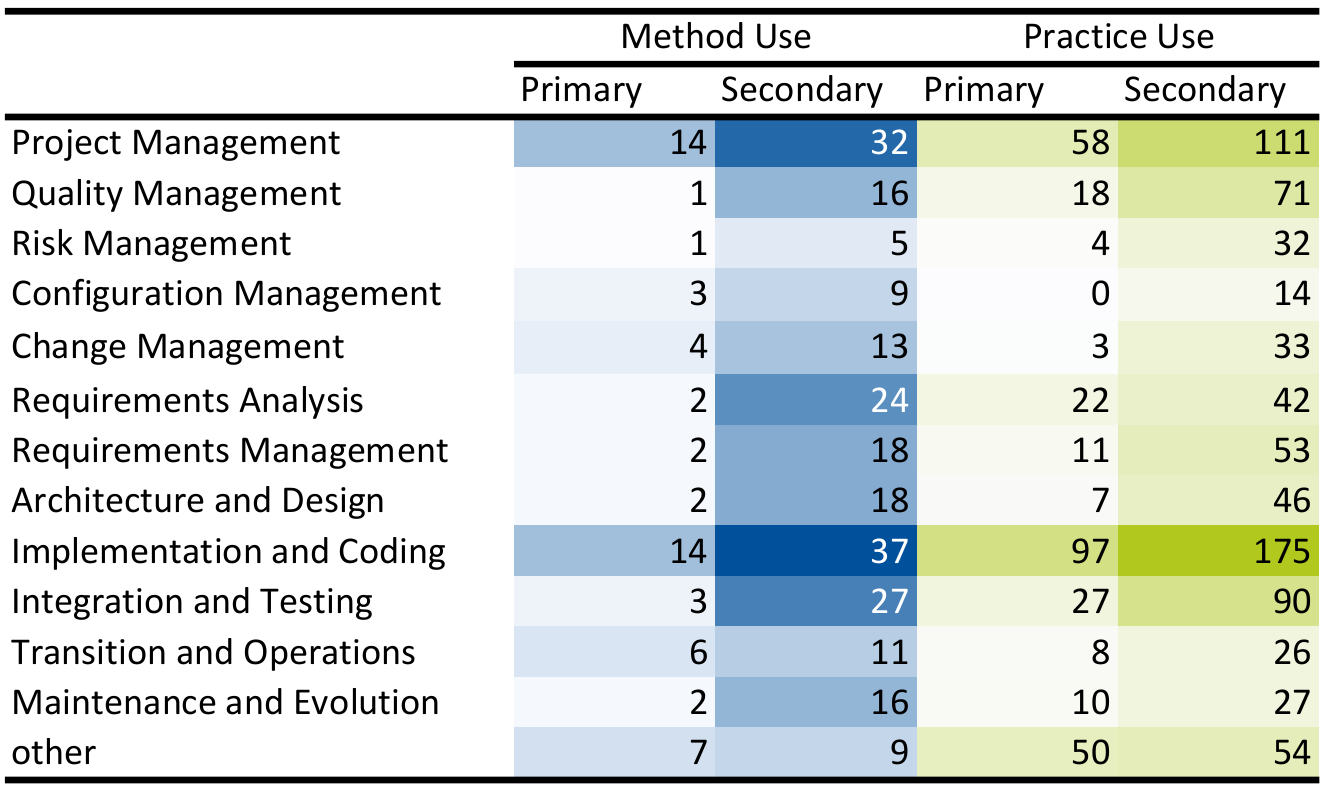}
	\caption{Total use of methods and practices for the different project disciplines (ordered by the total use of methods and practices). As each participant can assign the project discipline to different methods (independent of the rank), we obtain numbers higher than the total number of participants.}
	\label{fig:total-disciplines}
\end{figure}

\subsubsection{Use of the Top-3 Methods}
To answer RQ2, we analyze for which project disciplines the most frequently mentioned methods and practices are used. We focus on the top-3 methods and practices. 

Figure~\ref{fig:methoduse} visualizes for which project disciplines the top-3 methods \textit{Scrum} (see Figure~\ref{fig:scrum}), \textit{DevOps} (see Figure~\ref{fig:devops}), and \textit{Kanban} (see Figure~\ref{fig:kanban}) are used. 

\paragraph{Scrum}
In total, 16 respondents reported on using \textit{Scrum} (Figure~\ref{fig:total-use}). The use of \textit{Scrum} per project discipline is visualized in Figure~\ref{fig:scrum}. \textit{Scrum} is mainly used for \textit{Implementation and Coding} (13 responses), \textit{Project Management} (11 responses), and \textit{Requirements Analysis} (10 responses). Notably, all respondents that reported on using \textit{Scrum} for \textit{Requirements Analysis} mentioned it as secondary use only. In terms of the primary use, \textit{Scrum} is mainly used for \textit{Implementation and Coding} (7 responses), followed by \textit{Project Management} (3 responses) and \textit{Transition and Operations} (3 responses).

\paragraph{DevOps}
\textit{DevOps} are used by twelve respondents in our pilot study (Figure~\ref{fig:total-use}). The use of \textit{DevOps} is visualized in Figure~\ref{fig:devops}. \textit{DevOps} are mainly used for \textit{Integration and Testing} (6 responses), followed by \textit{Implementation and Coding} (4 responses), and \textit{Transition and Operations} (4 responses). The primary use of \textit{DevOps} is very manifold. Two respondents each reported on primarily using \textit{DevOps} for \textit{Implementation and Coding}, \textit{Transition and Operation}, \textit{Project Management}, and other. 

\paragraph{Kanban}
Twelve respondents reported on using \textit{Kanban} in the development process (Figure~\ref{fig:total-use}). Figure~\ref{fig:kanban} presents the use of \textit{Kanban} in the different project disciplines. \textit{Kanban} is mainly used for \textit{Project Management} (9 responses). However, these nine mentions reported on a primary use of \textit{Kanban} for \textit{Project Management} in two cases only. Besides, \textit{Kanban} is used for \textit{Implementation and Coding} (5 responses) and the \textit{Requirements Management} (5 responses). Focusing on the primary use, \textit{Kanban} was mentioned two times each for \textit{Project Management}, \textit{Implementation and Coding}, and other. 

\begin{finding}
	\noindent
		\textbf{Finding 3:} The use of Scrum, DevOps, and Kanban in the development process is very diverse. \\
		\textbf{Finding 4a:} Despite its nature as project management framework, Scrum is mainly (primarily and secondarily) used for the implementation and coding.\\
		\textbf{Finding 4b:} DevOps, which are originally meant to be used for operations, and Kanban, which is first and foremost a framework for work organization, are both primarily used for a number of different project disciplines.
\end{finding} 

\subsubsection{Use of the Top-3 Practices}
Besides the methods, Figure~\ref{fig:practiceuse} visualizes for which project disciplines the top-3 practices \textit{Code Reviews} (see Figure~\ref{fig:codereviews}), \textit{Daily Stand-up Meetings} (see Figure~\ref{fig:dailies}), and \textit{Coding Standards} (see Figure~\ref{fig:codingstandards}) are used. 

\paragraph{Code Reviews}
In our pilot study, 21 respondents reported on using \textit{Code Reviews} in the development process (Figure~\ref{fig:total-use}). Most of them use \textit{Code Reviews} for \textit{Implementation and Coding} (15 responses), \textit{Quality Management} (8 responses), and \textit{Integration and Testing} (6 responses). Notably, all but one respondent that reported on using \textit{Code Reviews} for \textit{Integration and Testing} mentioned it as secondary use only. Considering the primary use, \textit{Code Reviews} are mainly used for \textit{Implementation and Coding} (9 responses), followed by \textit{Quality Management} (5 responses). 

\paragraph{Daily Stand-up Meetings}
\textit{Daily Stand-up Meetings} are used by 21 respondents of our pilot study (Figure~\ref{fig:total-use}). Most of them use \textit{Daily Stand-up Meetings} for \textit{Project Management} (13 responses), followed by \textit{Implementation and Coding} (11 responses). However, only five respondents reported on using \textit{Daily Stand-up Meetings} mainly for \textit{Project Management}. The second most frequent use of \textit{Daily Stand-up Meetings} is reported for \textit{Implementation and Coding}, with a majority of ten participants stating to use it primarily for this discipline. All other disciplines are reported to be supported by \textit{Daily Stand-up Meetings} at most three times. 

\paragraph{Coding Standards}
Twenty respondents of our pilot study use \textit{Coding Standards} (Figure~\ref{fig:total-use}). The majority of them use \textit{Coding Standards} to support \textit{Implementation and Coding} (13 responses). The second most frequent use is reported for \textit{Quality Management}, with five mentions only. Considering the primary use of \textit{Coding Standards}, most frequently we find \textit{Implementation and Coding} (8 responses), again followed by \textit{Quality Management} (2 responses), as well as \textit{Integration and Testing} (2 responses). 

\begin{finding}
\textbf{Finding 5:} Compared to the methods, we observe a more stringent use of the practices for project disciplines.  \\
		\textbf{Finding 6a:} Code Reviews (which is a practice for Quality Management first of all)  as well as Coding Standards are mainly used for the Implementation and Coding. \\
		\textbf{Finding 6b:} Although they are meant to allow the synchronization of the whole team by pointing to finished or ongoing tasks as well as to issues, Daily Stand-up Meetings are most frequently primarily used for Implementation and Coding.
\end{finding} 

\subsection{Supported Project Disciplines}
Based on the selection of methods and practices used by more than ten participants (Figure~\ref{fig:total-use}), we analyze for which project disciplines the methods and practices are used. 

Figure~\ref{fig:total-disciplines} summarizes the number of methods and practices that are stated to be primarily (first two columns, colored in blue) or secondarily (last two columns, colored in green) used for a specific project discipline. For example, in the case of \textit{Maintenance and Evolution}, we find two mentions of a method primarily used for this discipline (this can be one participant ranking this discipline on first place for two methods, or two different participants), and in total 16 mentions of the discipline on another rank.

As evident from Figure~\ref{fig:total-disciplines}, \textit{Implementation and Coding} is the most frequently mentioned project discipline (in total 323 mentions), followed by \textit{Project Management} (215 mentions), \textit{Integration and Testing} (147 mentions), the ``\textit{other}'' option (120 mentions), and \textit{Quality Management} (106 mentions). These four disciplines (plus the ``other'' option) are each mentioned more than 100 times. Given the few mentions of other project disciplines, we do not present detailed results for these disciplines due to the limited informative value.

\begin{figure*}[!t]
	\centering
	\begin{subfigure}[b]{0.49\textwidth}
		\centering
		\includegraphics[width=\textwidth]{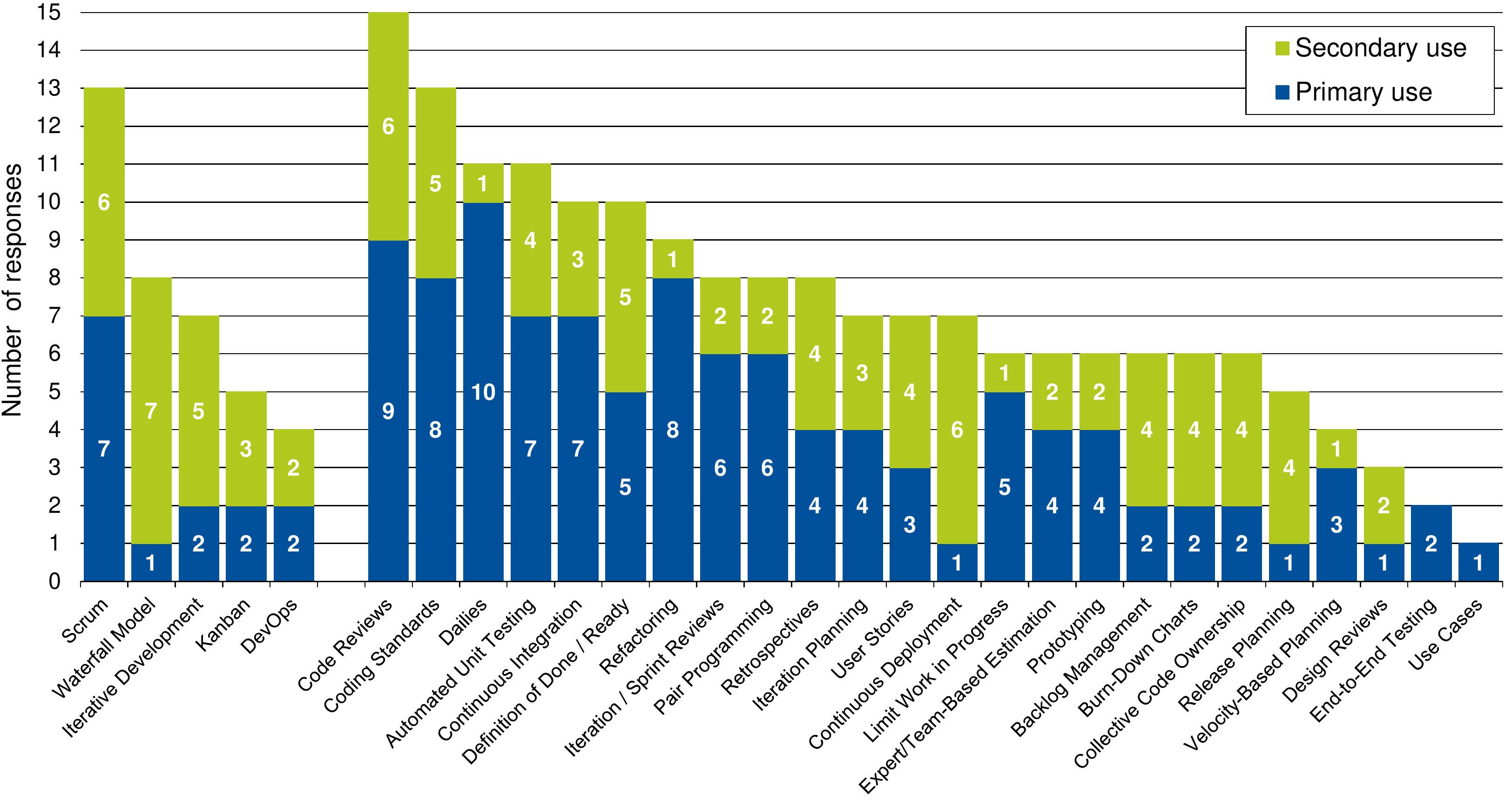}
		\caption{Methods and practices used for Implementation and Coding}
		\label{fig:disc-implcod}
	\end{subfigure}
	\hfill
	\begin{subfigure}[b]{0.49\textwidth}
		\centering
		\includegraphics[width=\textwidth]{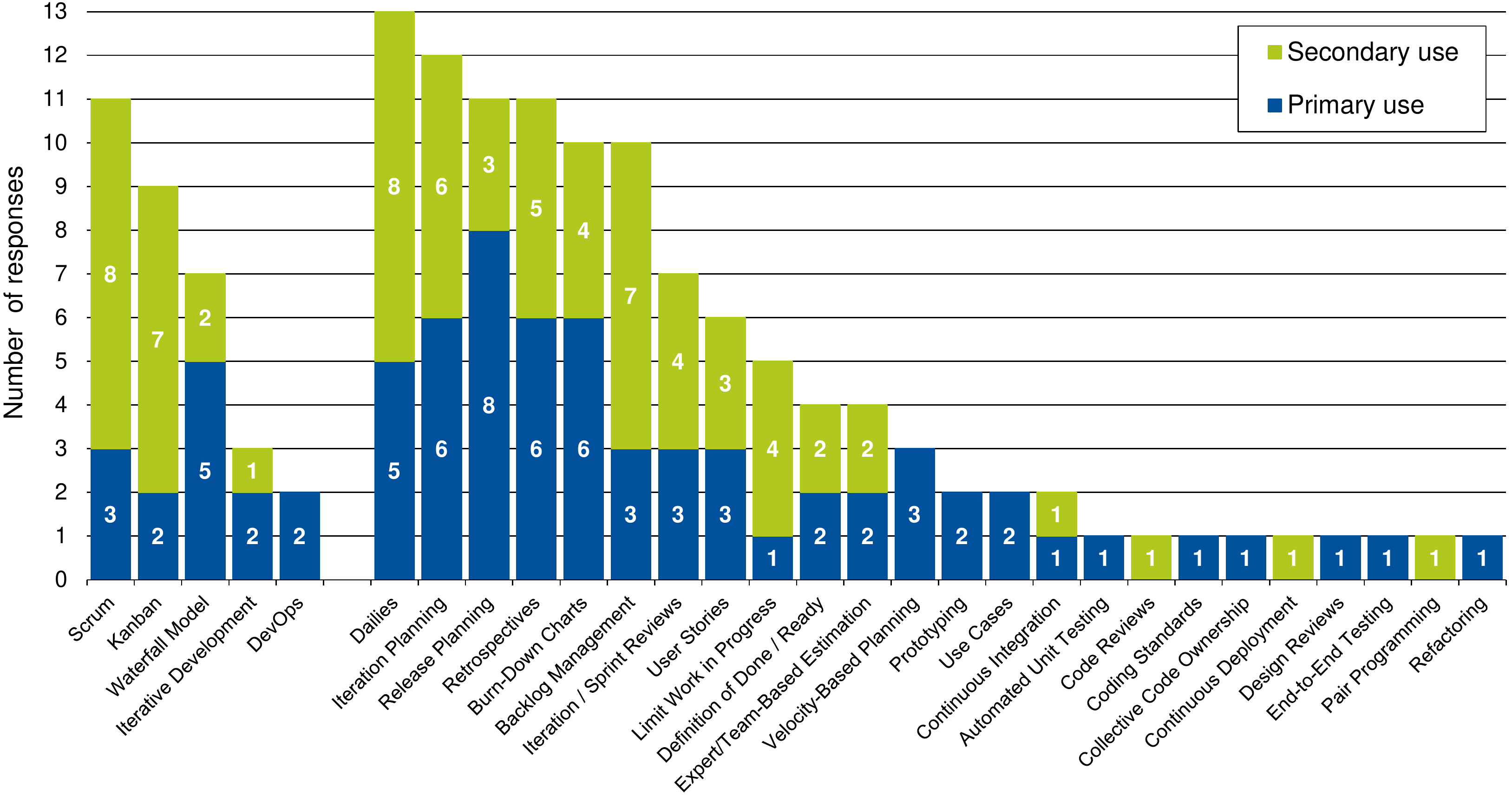}
		\caption{Methods and practices used for Project Management}
		\label{fig:disc-pm}
	\end{subfigure}
	\hfill
	\begin{subfigure}[b]{0.49\textwidth}
		\centering
		\includegraphics[width=\textwidth]{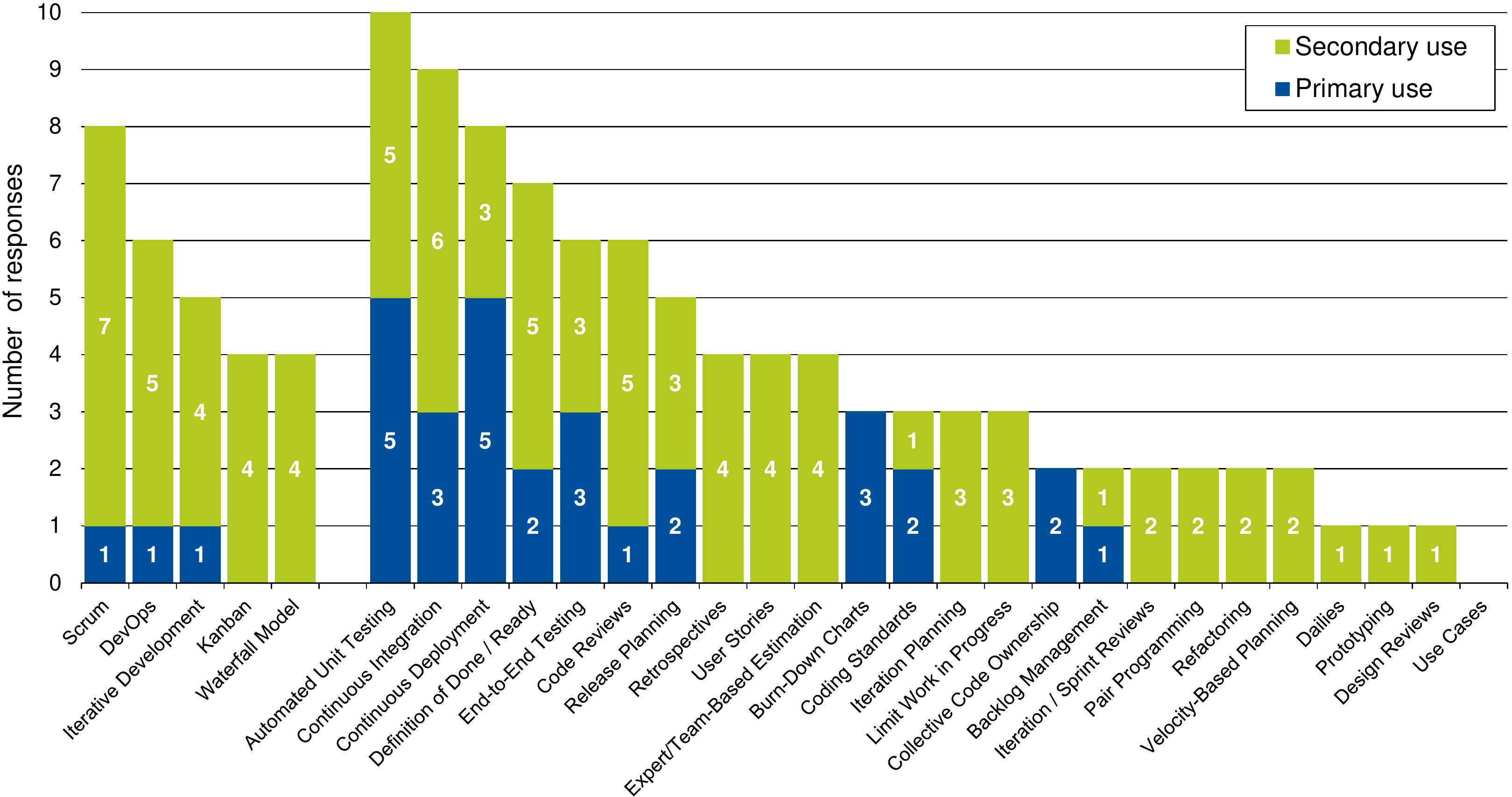}
		\caption{Methods and practices used for Integration and Testing}
		\label{fig:disc-inttest}
	\end{subfigure}
	\hfill
	\begin{subfigure}[b]{0.49\textwidth}
		\centering
		\includegraphics[width=\textwidth]{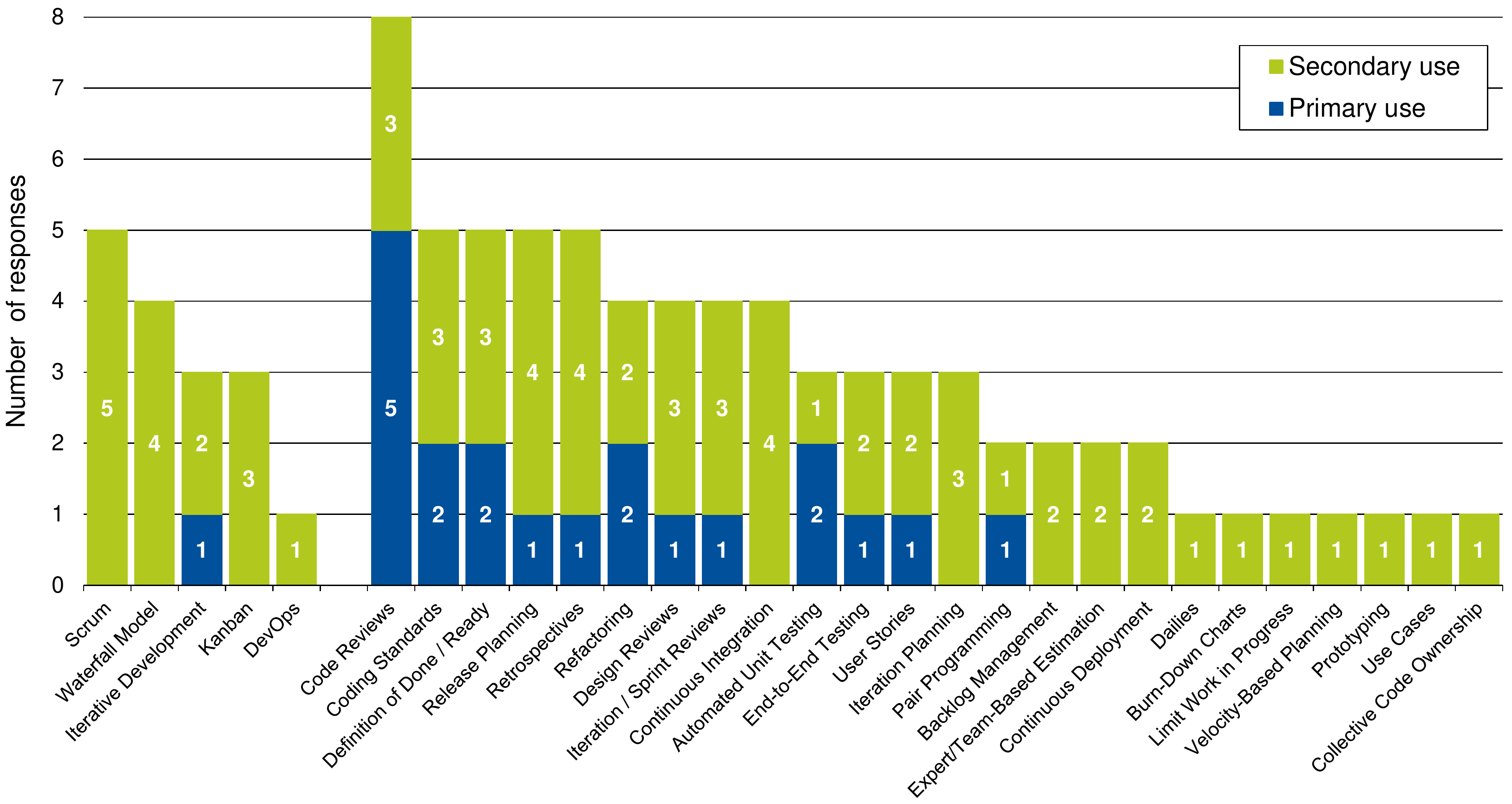}
		\caption{Methods and practices used for Quality Management}
		\label{fig:disc-qm}
	\end{subfigure}
	\caption{Overview of methods and practices used for the four most frequently mentioned project disciplines Implementation and Coding (a), Project Management (b), Integration and Testing (c), and Quality Management (d)}
	\label{fig:disciplines}
\end{figure*}

In a next step, for each of the four frequently mentioned project disciplines, we analyze which methods and practices are used primarily and secondarily. The results are summarized in Figure~\ref{fig:disciplines}, differentiating between the primary use of a method/practice for the respective project discipline (blue bars) and the secondary use (green bars).

\paragraph{Implementation and Coding}
Figure~\ref{fig:disc-implcod} visualizes the methods and practices used for \textit{Implementation and Coding}. As evident from Figure~\ref{fig:disc-implcod}, with in total 13 mentions, \textit{Scrum} is the most frequently used method for \textit{Implementation and Coding}, with seven respondents stating to use \textit{Scrum} primarily for this discipline. The \textit{Classic Waterfall Model} and the other three methods (\textit{Iterative Development}, \textit{Kanban}, and \textit{DevOps}) are mainly used secondarily for \textit{Implementation and Coding}. \textit{Code Reviews} (15 responses) are the most frequently mentioned practice for \textit{Implementation and Coding}, followed by \textit{Coding Standards} (13 responses) and \textit{Automated Unit Testing} (11 responses). 

Considering the primary use only, \textit{Scrum} is the most frequently mentioned method (7 responses), and \textit{Daily Stand-up Meetings} are the most frequently mentioned practice (10 responses), followed by \textit{Code Reviews} (9 responses), \textit{Coding Standards} (8 responses), and \textit{Refactoring} (8 responses). 

\paragraph{Project Management}
Figure~\ref{fig:disc-pm} visualizes the methods and practices used for the \textit{Project Management}. As evident from Figure~\ref{fig:disc-pm}, with a total of eleven mentions, \textit{Scrum} is the most frequently used method for \textit{Project Management}, with three respondents stating to use \textit{Scrum} primarily for this discipline. \textit{Daily Stand-up Meetings} are the most frequently mentioned practice for \textit{Project Management} (13 responses), followed by \textit{Iteration Planning} (12 responses), \textit{Release Planning} (11 responses), and \textit{Retrospectives }(11 responses).  

Considering the primary use only, the \textit{Classic Waterfall Model} is the most frequently mentioned method (5 mentions), and \textit{Release Planning} is the most frequently mentioned practice (8 mentions), followed by \textit{Iteration Planning}, \textit{Retrospectives}, and \textit{Burn-down Charts} (6 responses each).

\paragraph{Integration and Testing}
Figure~\ref{fig:disc-inttest} visualizes the methods and practices used for the \textit{Integration and Testing}. As evident from Figure~\ref{fig:disc-inttest}, with in total eight mentions, \textit{Scrum} is the most frequently used method for \textit{Integration and Testing} with only one respondent stating to use \textit{Scrum} primarily for this discipline. However, all five methods considered in our analysis are mostly secondarily used for \textit{Integration and Testing}. For \textit{Scrum}, \textit{DevOps}, and \textit{Iterative Development}, only one participant (for each method) stated to use the respective method primarily for \textit{Integration and Testing}. With ten mentions, \textit{Automated Unit Testing} is the most frequently mentioned practice to be used for \textit{Integration and Testing}, followed by \textit{Continuous Integration} (9 responses), and \textit{Continuous Deployment} (8 responses).

Considering the primary use only, the project discipline \textit{Integration and Testing} is only rarely covered by the methods and practices used in our study. For the five methods, we find a total of three mentions as primary use for this discipline. For the practices, we see some more mentions, with \textit{Automated Unit Testing} and \textit{Continuous Deployment} both being mentioned to be primarily used for \textit{Integration and Testing} five times, followed by \textit{Continuous Integration}, \textit{End-to-End Testing}, and \textit{Burn-down Charts} which are each mentioned three times.
 
\paragraph{Quality Management}
Figure~\ref{fig:disc-qm} visualizes the methods and practices used for the quality management. As evident from Figure~\ref{fig:disc-qm}, with a total of five mentions, \textit{Scrum} is the most used method for \textit{Quality Management}, but all five respondents using \textit{Scrum} for \textit{Quality Management} state to use it only secondarily for this discipline. \textit{Code Reviews} are the most frequently mentioned practice for \textit{Quality Management} (8 responses), followed by \textit{Release Planning}, \textit{Retrospectices}, \textit{Definition of Done/Ready}, and \textit{Coding Standards} which are each mentioned five times. 

Considering the primary use only, \textit{Iterative Development} is the only method considered in our study that is reported to be primarily used for \textit{Quality Management}---and even this method is only mentioned once in this context. \textit{Code Reviews} are the most frequently mentioned practice to support the \textit{Quality Management} (5 responses). In total, of the 16 mentions of methods used for the \textit{Quality Management}, only one reports on a primary use. In case of the practices, we have 71 mentions, with 20 mentions reporting on a primary use.

\begin{finding}
	\textbf{Finding 7:} We observe a huge variety in the (primary) use of methods and practices for the different project disciplines. \\
		\textbf{Finding 8a:} Considering the five methods, we often do not see a clear support of a method for the respective discipline. \\
		\textbf{Finding 8b:} While we observe a wide coverage of primary uses for the disciplines Implementation and Coding, Project Management, Integration and Testing, as well as Quality Management are mainly covered secondarily by the methods and practices considered in our pilot study.
\end{finding}

\section{Discussion}\label{sec:discussion}
Based on our results, we conclude this paper by answering the three research questions, discussing our results, and presenting the threats to validity.

\subsection{Answering the Research Questions}
Summarizing the results, we can answer the research questions posed in Section~\ref{sec:rqs} as follows:\\
\textbf{Answer to RQ1:} The participants mainly reported on using \textit{Scrum}, \textit{Kanban}, and \textit{DevOps}. The most frequently used practices are \textit{Daily Stand-up Meetings}, \textit{Code Reviews}, and \textit{Coding Standards} (Figure~\ref{fig:total-use}). \\
\textbf{Answer to RQ2:} The methods and practices are mainly used for \textit{Implementation and Coding}, followed by the \textit{Project Management}, \textit{Integration and Testing}, as well as \textit{Quality Management}. \textit{Configuration Management} and \textit{Risk Management} are only rarely reported on being supported by the methods and practices covered by our study (Figure~\ref{fig:total-disciplines}).\\
\textbf{Answer to RQ3:} Looking at the four project disciplines mentioned more than 100 times in our pilot study, \textit{Scrum} is the most frequently used method for all of them. \textit{Implementation and Coding} is mainly supported by \textit{Code Reviews} and \textit{Coding Standards}, which are also the most frequently used practices for \textit{Quality Management}. The \textit{Project Management} is mainly supported by \textit{Daily Stand-up Meetings} and \textit{Iteration Planning}, \textit{Integration and Testing} is supported by \textit{Automated Unit Testing} and \textit{Continuous Integration}. 

\subsection{Interpretation}
As software projects are very diverse and operated in different contexts, software processes are rather unique than mainstream, and adjusted to the project's and team's needs~\cite{tell2019hybrid}. So far, research has focused on statistical methods to construct a hybrid development method~\cite{tell2020towards} and on context factors that influence the choice of specific methods~\cite{klunder2020determining}. In addition, most hybrid development processes are devised as part of a software process improvement initiative, and based on experience~\cite{klunder2019catching}. 

However, research has not yet considered the fact that a holistic development process needs to cover the whole software lifecycle---from the planning in the beginning to the maintenance and evolution in the end---and that this fact can be used when devising a hybrid method. Therefore, when constructing a development method, project managers and/or process engineers should analyze whether an implemented or proposed development process with the methods and practices used addresses all phases of the software project or not. As a first step to reach this goal, a mapping of different methods and practices to the different phases of the project is required. Therefore, in this paper, we investigated for which project disciplines specific methods and practices are used. Based on a rather small sample consisting of 27 data points reporting on 1,257 pairs of method/practice and project discipline, we observe two remarkable things:

(1) The methods and practices considered in our study rarely shape the whole software lifecycle (Figure~\ref{fig:total-disciplines}). 

(2) Several methods and practices are not used as they are meant to. For example, \textit{Scrum} as a project management framework is only used as such (and hence supporting the \textit{Project Management}) in three cases, whereas it is used primarily for \textit{Implementation and Coding} in seven cases (Figure~\ref{fig:scrum}).

Besides the threats to validity we discuss in Section~\ref{sec:threats}, there are two possible explanations for this (mis-)use of methods and practices as observed in the case of \textit{Scrum}. First, it is possible that the original method, e.g., \textit{Scrum}, is still considered as \textit{Scrum} although it is adjusted to fit the team's need and not even the \textit{ScrumBut phenomenon}\footnote{ScrumBut is an adjustment of the original Scrum framework to solve a dysfunction in a team on symptom level without finding and solving the origin of the problem. Further information can be found at the Scrum.org website: \url{https://www.scrum.org/resources/what-scrumbut}.}(\cite{eloranta2016exploring}) anymore. This goes along with Tell et al.'s~\cite{tell2020towards} finding 643 widely distributed variations of \textit{Scrum} which are all considered \textit{Scrum}. It is not unlikely that some of these variations also contain practices (and other methods) contributing to the \textit{Implementation and Coding}. These issues raise the question of what is the value of talking about the use of \textit{Scrum}---in research and practice---if no-one does it by the book? And, as long as everybody has an own idea of what \textit{Scrum} looks like, how it should be used, and what it should be used for, what is the value of stating to use it? Is it just used as a brand and a marketing tool as stated by Hohl et al.~\cite{hohl2018back}? 

Based on our results, we are not able to answer these questions. Nevertheless, our results point to the missing baseline of what \textit{Scrum} really is about. Consequently, future research on software processes (and how they are shaped with methods and practices) should carefully look at the mental model of the methods and practices to create a valid baseline. In addition, practitioners should be aware of three things: (1) whether they deviate from a specific method/practice, (2) how they do it, and (3) why. As long as this awareness is present and the adaptation of \textit{Scrum} is suitable and based on experience~\cite{klunder2019catching}, it is worth a thought to leave the beaten path and to adjust \textit{Scrum} to the specific needs~\cite{mortada2020software,diebold2015practitioners}.

This line of thought is also valid for \textit{DevOps} and \textit{Kanban}. \textit{DevOps}, which are originally meant to be used for the operations part of a project, are more or less used for everything (Finding 4b). \textit{Kanban} as a work organization technique with its basic idea of limiting the work-in-progress for work-flow organization is primarily used for a number of different project disciplines. How can the idea of limiting work-in-progress be helpful for the implementation and coding?

As \textit{Scrum}, \textit{DevOps}, and \textit{Kanban} are seen as rather generic methods, they are applied in a variety of project disciplines: If you have a hammer, every problem looks like a nail. That is, when building a development process, we---researchers and practitioners---should first and foremost go back to the original idea behind a method or a practice, before using it in a way and in a context in which it cannot work. Of course, you can also sink a screw into the wall with a hammer, but isn't it more practical to use other tools?  Most of the methods and practices are not as generic as they are seen (e.g., \textit{Scrum} focuses on Project Management, \textit{Kanban} on work organization, etc.). 

However, this line of thought can also be reversed: Probably, practitioners have an established and well functioning development process and just need a ``fancy'' name. This again contributes to Hohl et al.'s findings~\cite{hohl2018back} of seeing \textit{Scrum} as a marketing tool. If your process does not have a name highlighting that it is based on one of these methods on which every process is based, it will not function. 

Nevertheless, there is a second possible explanation for the wide variety of use cases of these generic methods: It is also possible that the practitioners do not know exactly why a specific method or practice is used. This increases the risk of a development team rejecting the development process because the rationale behind using specific methods and practices does not come clear.

In addition, it appears that it is easier to address some dominant project disciplines such as \textit{Implementation and Coding} or \textit{Project Management} with methods and practices, whereas other disciplines such as the \textit{Configuration Management} are hardly covered by ``standard'' methods and practices. Consequently, by just using the well-known methods and practices, it is almost impossible to address the whole software project. When devising a hybrid development method, it does not suffice to implement the widely distributed methods and practices because this would lead to uncovered phases during the project lifetime.

\subsection{Threats to Validity}\label{sec:threats}
Despite the implemented validity procedures we described in Section~\ref{sec:validity}, the results of our study are subject to some limitations and threats to validity. In this section, we discuss the threats to validity according to Wohlin et al.~\cite{wohlin2012experimentation}.

\paragraph{Construct validity}
The results of our pilot study are based on an online questionnaire. Therefore, the options provided in the multiple-choice questions, in particular the lists of methods and practices, might have been incomplete. In addition, we cannot guarantee that participants did not misunderstand some questions, and that everybody (researchers and participants) has the same idea of what, e.g., \textit{Scrum} looks like. To reduce the risk of misinterpretations, two independent researchers reviewed the questionnaire, and we published the questionnaire in three different languages to avoid misunderstandings caused by language issues. 

In addition, the convenience sampling strategy may have led to participants not belonging to the target audience. However, executing the survey requires specific knowledge on the meaning of methods, practices, and project disciplines. As all answers but one provide meaningful and realistic insights, we are confident that the participants reflect the target audience. 

\paragraph{Internal validity} 
Maturation and mortality are two important threats to validity for a survey. In particular, the time required to complete the questionnaire is crucial. Too many questions can negatively affect respondents probably leading to abort. This has to be taken into account when developing the survey. We mitigated this threat by keeping the survey as short as possible (although this introduced the threat of incomplete answers).

In general, survey results reflect the perspective and opinion of the participants, so the results cannot be considered objective.

Prior to the analysis, we cleaned the data following the procedure described in Section~\ref{sec:data-analysis-procedures}. The data analysis also followed a stringent procedure performed by two researchers and was carefully reviewed by the other authors of this paper.

\paragraph{Conclusion Validity}
All analyses we performed in this paper are descriptive. Due to the small sample size of 27 data points, it was not possible to perform any statistical tests that might have strengthened the reliability of our results. However, this pilot study was meant to provide a first picture on the use of methods and practices throughout the development process, and to motivate future research. Therefore, the results should be interpreted carefully, as they are based on insights from a few companies only. Further research is necessary to reduce these risks. 

\paragraph{External validity}
We invited companies of different sizes to participate in the survey. However, the results do not include participants of all targeted company sizes. In order to be able to make generalized statements, the survey must be replicated with more participants and from differently sized companies.

\section{Conclusion}\label{sec:conclusion}
Using a pilot study with 27 participants, we present initial results on the use of methods and practices in the different phases of a software project. Based on our results, we conclude three key findings: (1) Methods are often not used as they are meant to be used, (2) practices are more consistently used than methods, and (3) parts of the development process are hardly covered by widely distributed methods and practices.

For example, \textit{Scrum} and \textit{Kanban} as Project Management frameworks are more or less used throughout the whole software lifecycle, including the implementation and coding. This raises the questions of how, for instance, a project management framework such as \textit{Scrum}---adjusted in a way such that it is still considered \textit{Scrum}---can contribute to the technical discipline \textit{Implementation and Coding}, and whether the use of Scrum in this phase can be meaningful at all. It appears to be state of the art to use a rather generic framework for almost everything throughout the project. In this case, research and practice should go back to the roots and the basic ideas behind the methods and practices to derive meaningful and well-working software processes. 

In addition, our results indicate that there is a wide variety of methods and practices for project disciplines such as \textit{Implementation and Coding}, whereas it appears to be more difficult to address other disciplines such as \textit{Configuration Management}. Nevertheless, future research is required to analyze whether these results are caused by the pre-selection of methods and practices we used in our study, or, if we observe similar results based on a larger dataset. In addition, further studies are required to increase the reliability of our results. For example, a thorough comparison with the original literature on the methods and practices and how they should be used is still missing. In addition, interview studies would be meaningful to go into more detail in specific cases. 

Summarizing, our results indicate that research and practice should put emphasis on the original ideas of a method or practice, and scrutinize whether it is meaningful or not to integrate specific methods or practices in the development process. It is not helpful at all to consider generic methods as being able to solve every problem.

\section*{Acknowledgments}
We thank all the study participants for the provided insights on their development process. 
\bibliographystyle{IEEEtran}
\bibliography{IEEEabrv,references}
\end{document}